# Asymptotically stable phase synchronization revealed by autoregressive circle maps

F.R. Drepper

Zentrallabor für Elektronik
Forschungszentrum Jülich GmbH
52425 Jülich, Germany
f.drepper@fz-juelich.de

Jülich, 25.07. 2000

A new type of nonlinear time series analysis is introduced, based on phases, which are defined as polar angles in spaces spanned by a finite number of delayed coordinates. A canonical choice of the polar axis and a related implicit estimation scheme for the potentially underlying autoregressive circle map (next phase map) guarantee the invertibility of reconstructed phase space trajectories to the original coordinates. The resulting Fourier approximated, Invertibility enforcing Phase Space map (FIPS map) is well suited to detect conditional asymptotic stability of coupled phases. This rather general synchronization criterion unites two existing generalisations of the old concept and can successfully be applied e.g. to phases obtained from ECG and airflow recordings characterizing cardio-respiratory interaction.

The relevance of deterministic dynamics is not limited to situations, where the rules of change can be derived by a mechanistic (deductive) approach. The alternative (inductive) approach is aimed at the estimation of phenomenological dynamic models, based on the use of a sufficient amount of empirical data. A common starting point in modern, nonlinear time series analysis [1-3] is the assumption that a state space reconstruction based on a finite number of succeeding samples (coordinates) represents an embedding of the underlying deterministic dynamics, i.e. reconstruction and original dynamics are expected to be topologically equivalent. The latter feature implies the existence of a diffeomorphic map (smooth map with smooth inverse) linking reconstructed attractors to the ones of the underlying dynamics.

One of the major problems in nonlinear time series analysis is to find a sufficiently general class of autoregressive models for which the parameter estimation (regression) problem is feasible. One promising solution is the introduction of angle or phase type variables prior to the estimation step. Due to the periodic character of phase variables this additional transformation allows the use of finite Fourier series to approximate the phase space dynamics. (It is important to note that the notion "phase space" is often used as a synonym for state space, in contrast to the present more specific meaning based on angles.) However the transformation of time series to the well-known Hilbert phase [4-7] is an irreversible step which excludes e.g. to study the transformation properties of Hilbert phase dynamics to the original coordinates or to different phase type variables. As a related feature, Hilbert phases evidence a trivial long-range autocorrelation which is difficult to distinguish from an underlying determinism.

A basic idea of the present study is to use the ambiguity of the phase definition to select a special class of (canonical) phases, for which the smoothness of the backtransformation of reconstructed phase space trajectories can be guaranteed. This guaranty is enforced by a related implicit estimation approach to auto-regressive circle maps (next phase maps) which express the a priori correlation of successive phases and the potentially underlying low dimensional dynamics. This way Fourier approximated, Invertibility enforcing Phase Space maps (FIPS maps) may be seen as a promising alternative to existing global approximation methods in nonlinear time series analysis like radial basis function approximation [8,9]. The main advantage however turns out in the multivariate case where FIPS maps can be used to analyze synchronization phenomena, including those with instationary amplitudes.

The old concept [10] of synchronization has recently received two important extensions [5,6,11-15] opening the notion to non-identical and aperiodic coupled oscillators: generalized synchronization of unidirectionally coupled oscillators [12-15] and phase synchronization [5,6,11]. The central notion of generalized synchronization is an asymptotically stable map [12-16] (Afraimovich map) which relates simultaneous states of the two oscillators (subsystems). The detection of generalized synchronization in the situation of time series analysis has so far been focused on continuity [15] or predictability [17] properties of the Afraimovich map. The present study uses the universal and efficient approximation features of FIPS maps to introduce the asymptotic stability criterion to time series analysis. In spite of the smoothness of the dynamics expressed by (finite order) FIPS maps this criterion is very flexible concerning the properties of their asymptotically stable maps. Thus asymptotic stability of coupled FIPS map reconstructions turns out as a unifying criterion for a wealth of synchronization phenomena including generalized synchronization in unidirectionally coupled systems and the whole family of (n:m) phase synchronization.

The new criterion turns out to contribute to the answer of a recently disputed question in theoretical physiology: whether cardio-respiratory interaction leads to synchronization [6,18]. Cardio-respiratory interaction, whose scientific record dates back to 1849 [19], refers to the fact that (preferentially in relaxed state) heartbeat intervals have the tendency to become shorter during inspiration and longer during expiration [20,21]. Numerous studies can be taken as evidence that the main part of this interaction is not a mechanical one but a neural one linking cardio-pulmonary and respiratory centers in the brain stem [22-23] as well as periferal baro-receptors [22,24]. Prior to their application to the empirical example the new concepts are tested on a 2-dimensional, time discrete model system.

## Canonical phases

The first step of a FIPS map analysis introduces polar angles into a delay co-ordinate based (primary) reconstruction space of an oscillating signal $\{X_n | n=0,1,...,N+1\}$ sampled at times $\{t_n | n=0,1,...,N+1\}$. The sampling rate is assumed to lie in the range from about 3 to 8 samples per lowest (typical) cycle length.

Furthermore it is assumed that time averages of the signal are close to zero when their range extends over periods which are large in comparison to the largest typical (recurrent) cycle length. To ensure these two features, a low and/or high pass filter might have to be applied in a preparatory step. In contrast to the well-known Hilbert phase [4-7] a low dimensional phase definition space is used. For notional simplicity its dimension will be restricted to a maximum of three. Within this space two linear functions $L_X$ and $L_Y$ are chosen which express projections along two different orthogonal directions defining the polar axis. Based on these two functions, signal $X$ is transformed to a set of phases and radii, $\{\varphi_n, r_n | n=1,...,N\}$, where

$$\varphi_n = \arctan 2\left(L_x(X_{n-1}, X_n, X_{n+1}),\ L_y(X_{n-1}, X_n, X_{n+1})\right) \quad (1a)$$

$$r_n = \sqrt{L_x^2(X_{n-1}, X_n, X_{n+1}) + L_y^2(X_{n-1}, X_n, X_{n+1})}. \quad (1b)$$

The bivariate function $\arctan 2(x, y)$ extends the range of $\arctan(y/x)$ to the full circle from $-\pi$ to $\pi$. Equations (1a) and (1b) represent a transformation $\Re^{N+2} \to \Re^{2N}$. As an intermediate step towards its inversion, transformation (1) can be brought into the following implicit form

$$\begin{aligned} L_x(X_{n-1}, X_n, X_{n+1}) &= r_n \cos\varphi_n \\ L_y(X_{n-1}, X_n, X_{n+1}) &= r_n \sin\varphi_n \end{aligned} \quad \text{for } n=1,2,...,N. \quad (2)$$

For given phases $\varphi_n$ and radii $r_n$, equation set (2) represents an overdetermined linear equation system for $X_0, ..., X_{N+1}$ with at least N-2 invertibility constraints on the left hand side of (2). These constraints impose restrictions on the set of possible combinations of succeeding values of $r_n$ and $\varphi_n$ for which the backtransformation to original coordinates $X_n$ is feasible. The $J \geq N-2$ constraints on the inhomogenous part of (2) can be expressed with the help of a set of linear independent solutions $\{H_j | j=1,...,J\}$ of the corresponding homogenous transposed equation system. These nontrivial solutions are assumed to span the so-called null space. The right hand side of (2) has to be orthogonal to all these solutions, i.e.

$$\sum_{n=1}^{N} \left[ H_{j,2n-1} \, r_n \cos\varphi_n \; + \; H_{j,2n} \, r_n \sin\varphi_n \right] \; = \; 0 \quad \text{for } (j=1,...,J). \qquad (3)$$

For canonical phases $L_X$ and $L_Y$ have to be chosen in such a way that there exists a complete set of linear independent solutions $\boldsymbol{H}_j$ for which each $\boldsymbol{H}_j$ has at most four nonzero components. Furthermore these nonzero components have to fulfill the property that every single equation of (3) relates only one pair of radii and corresponding angles. In the case of adjacent pairs, constraint set (3) assumes the simplified form

$$r_{j+1} \cos(\varphi_{j+1} - \beta) \; - \; r_j \sin(\varphi_j - \gamma) \; = \; 0 \quad \text{for } (j=1,...,J), \qquad (4)$$

where the constants $\beta$ and $\gamma$ can be expected to have no j dependence due to translational symmetry in time. The canonical form of constraint set (4) implies a separation of the constraints into restrictions on combinations of succeeding phases and ones on succeeding radii. If one of the angular functions becomes zero the other one has to be zero too. Once a permissible set of phases is obtained, (4) can be interpreted as recursion formula for the radii. Canonical phases do not automatically guarantee the diffeomorphic character of their backtransformation, however the simple (radius independent) form of the restriction on successive phases opens the way to define reconstructions of phases which are diffeomorphically related to the corresponding reconstructions of original coordinates.

In the case of a 2-dimensional phase definition space there exists a one parameter family of canonical phases, $L_x = \cos\alpha\, X_n - \sin\alpha\, X_{n+1}$ and $L_y = \sin\alpha\, X_n + \cos\alpha\, X_{n+1}$, leading to $J = N - 1$, $H_1 = (\tan\alpha, 1, -1, \tan\alpha, 0, ...,0)$, $H_2 = (0, 0, \tan\alpha, 1, -1, \tan\alpha, 0, ...,0)$, ... and to $\beta = \gamma = \alpha$. The examples of this study will use the most simple symmetric case $\alpha = 0$. Table 1 also lists two 3-dimensional cases: $L_x = (X_{n-1} - X_{n+1})/\sqrt{2}$, $L_y = (X_{n-1} \pm 2X_n + X_{n+1})/\sqrt{6}$, leading to $J = N - 2$ and $H_1 = (1, -\sqrt{3}, \pm 1, \pm\sqrt{3}, 0,...,0)$. The first one of these can be interpreted as finite dimensional analogon of the Hilbert phase. The last example is of particular interest for highly instationary data, because in this case both linear functions represent a high pass filter. All cases of Table 1 can be generalized by introducing a shift of the center (polar axis) along the space diagonal or by coarsening the time step length used for the definition of the primary reconstruction space. An example in place is $L_x = X_{n-1}$ and $L_y = X_{n+1}$. Such coarsening of the time scale might be useful when the sampling rate is higher than eight. For notational simplicity these cases will not be treated explicitly in the following.

**FIPS maps**

The next analytical step replaces the well known Takens type reconstruction [1,2] by a reconstruction based on delayed canonical phases [25]. A central problem of phase space reconstruction is the unavoidable correlation of successive phases due to their non-local in time relationship to the original coordinates. The following (phase type) homogenous function of degree zero evidences zero auto-correlation for a white noise signal X and is therefor particularly suited to act as the left hand side of a phenomenological regression model.

$$\frac{X_{n+2}}{r_n} = F(\varphi_n, \varphi_{n-1}, ...) \tag{5}$$

Since all functions of phases can be assumed to be periodic with period $2\pi$, the function F can be approximated efficiently by a finite Fourier series.

$$F(\varphi_n, \varphi_{n-1}, ...) = \sum_{k=0, l=-L, ...}^{k=K, l=L, ...} \left[ a_{k,l,...} \cos(k\varphi_n + l\varphi_{n-1} + ...) + b_{k,l,...} \sin(k\varphi_n + l\varphi_{n-1} + ...) \right] \tag{6}$$

Each pair of parameters $a_{k,l,...}$ and $b_{k,l,...}$ can be aggregated to a corresponding amplitude $A_{k,l,...} = \sqrt{a_{k,l,...}^2 + b_{k,l,...}^2}$. The zero autocorrelation property justifies the simplifying assumption of independently and identically distributed residuals of ansatz (5) and (6). Thus the additional assumption of Gaussian distributed residuals reduces the maximum likelihood estimation of the parameters to the standard problem known as multivariate linear least squares regression.

However for a white noise signal the left-hand side of (5) is known to be distributed according to a *t*-distribution with two degrees of freedom, a distribution which clearly deviates from the Gaussian. The study of the histograms of several examples (including the ones of the present study) suggests that a distribution with a Gaussian center and exponentially distributed tails (starting at about one standard deviation) represents a good choice. The maximum likelihood estimation based on such (or other selfconsistently chosen) distributions can easily be approximated with the help of an iterative weighted least square algorithm starting with the Gaussian case.

The estimates of the parameters of (6) can already be used to reconstruct a time series, however we prefer to express the left-hand side of (5) in terms of canonical phases. When using the two-dimensional phases, the left-hand side can directly be expressed in terms of phases $\varphi_{n+1}$ and $\varphi_n$. However in the three-dimensional cases the left hand side of (5) has to be extended by a second similar expression,

$$\frac{X_{n+2} \pm X_{n+1}}{r_n} = \frac{\sqrt{6}}{2}\left(\tan(\varphi_{n+1} - \tfrac{\pi}{3}) + \tan(\tfrac{\pi}{6})\right) \sin(\varphi_n - \gamma). \qquad (7)$$

All cases can be summarized in the general implicit form of the (univariate) FIPS map containing three constants $\beta, \gamma, \delta$ which differ for the different canonical phases as indicated in Table 1.

$$\left(\tan(\varphi_{n+1} - \beta) - \tan(\delta - \beta)\right) \sin(\varphi_n - \gamma) = G(\varphi_n, \varphi_{n-1}, ...) \qquad (8)$$

In analogy to (6) the phenomenological (FIPS map generating) function $G$ can be approximated efficiently by a finite Fourier series. As can easily be seen from (8), the finiteness of the estimated parameters guarantees the diffeomorphic character of transformation (1) by avoiding the singular combinations of successive phases, defined by the angular part of constraint set (4).

Due to their smooth invertibility, FIPS map reconstructions based on one of the canonical phases can be transformed in a topologically equivalent way to all other canonical phases. Furthermore in the case of the two-dimensional canonical phases the reconstructions for different $\alpha$ turn out to be identical.

Now the choice of the term canonical phase has become more obvious. The enlargement of the N-dimensional phase space by N additional, canonically conjugate radii has created the possibility to define canonical transformations $\Re^{2N} \to \Re^{2N}$ which preserve the form of the equations of motion: the recursion formula for the radii given in (4) as well as autoregressive circle map (8). Solving Eq. (8) for $\varphi_{n+1}$ leads to the explicit form of the FIPS map,

$$\varphi_{n+1} = \arctan 2\bigl(\sin(\varphi_n - \gamma), G(\varphi_n, \varphi_{n-1},...) + \tan(\delta - \beta)\sin(\varphi_n - \gamma)\bigr) + \beta, \quad (9)$$

where $G(\varphi_n, \varphi_{n-1},...)$ represents the estimated reconstruction generating function. Stochastic reconstructions are obtained by adding state independent white noise to the generating function. The standard deviations of the noise terms can be related to the standard errors of the corresponding linear regressions. (Table 2 uses Gaussian white noise with a standard deviation reduced by $1/2$).

FIPS maps can be generalized to the multivariate case characterized by additional signals and corresponding phases (preferentially of the same canonical type). For notational simplicity we assume that the FIPS map generating function $G$ is extended by one single phase $\{\psi_n | n = 1, \text{к}, N\}$. Thus finite Fourier series (6) describing the autonomous behavior has to be supplemented by further terms describing the cross impact (or mutual control) of the two oscillators: the open loop part depending exclusively on phase $\psi$ and the closed loop part containing the mixed terms. In the latter part at least two of the indices $k, l, ...$ are nonzero.

$$G(\varphi_n, \varphi_{n-1}, ...; \psi_n, \psi_{n-1}, ...) = F(\varphi_n, \varphi_{n-1}, ...) +$$

$$\sum_{k=0, l=-L, ...}^{k=K, l=L, ...} \left[ o_{k,l,...} \cos(k\psi_n + l\psi_{n-1} + ...) + p_{k,l,...} \sin(k\psi_n + l\psi_{n-1} + ...) \right] \quad (10)$$

$$+ \sum_{k=-K, l=-L, ...}^{k=K, l=L, ...} \left[ c_{k,l,...} \cos(k\varphi_n + l\psi_n + ...) + d_{k,l,...} \sin(k\varphi_n + l\psi_n + ...) \right]$$

The dynamics of $\psi$ are described by the analoga of (8) and (10) with exchanged roles of $\varphi$ and $\psi$. Cross impact amplitudes like $O_{k,l,...} = \sqrt{o_{k,l,...}^2 + p_{k,l,...}^2}$ are defined in analogy to their autonomous counterparts.

### Conditional asymptotic stability

The reconstruction of coupled bivariate dynamics can be used to execute a gedanken experiment designed to identify asymptotic stability [12] (AS) which can be examined with the help of conditional Liapunov exponents [14] (LE). The latter ones are based on the experiment that the dynamics of one of the two coupled oscillators (subsystems) is temporarily disturbed and the dynamics of the other oscillator is kept fixed corresponding to the unperturbed attractor. A negative (largest) conditional LE indicates that the (weakly) perturbed subsystem approaches the reference trajectory in such a way that the initial perturbation is forgotten. If this holds, the conditional LE quantifies the asymptotic exponential rate of convergence or AS. Efficient algorithms for normal and conditional LE make use of the Jacobi matrix [3] (linear expansion around the momentary state of autoregressive circle map (9) and (10)).

To make sure that the asymptotic stability of the perturbed subsystem is not a trivial one resulting either from a periodic attractor of an exclusively autonomous dynamics or from a fixed point, a second criterion is necessary which ensures an asymptotically noticeable cross impact from the second subsystem. One way to achieve this is to compare the conditional predictability with the corresponding unconditional (autonomous) one. This can be done either based on locally linear prediction [17], corresponding nearest neighbor statistics [26] or based on FIPS map reconstruction.

A second particularly attractive way in the present context is to use a significance test for nonzero cross impact amplitudes of the corresponding FIPS map. The sum of squares of the ratios of the constituents of open loop cross impact $O$ to their own estimation errors e.g. represents a useful statistic (for testing the presence of simultaneous cross impact).

$$\chi_O^2 = \sum_{k=1}^{K} \left[ \frac{o_{k,0,\ldots}^2}{\Delta o_{k,0,\ldots}^2} + \frac{p_{k,0,\ldots}^2}{\Delta p_{k,0,\ldots}^2} \right] \quad (12)$$

For ideal surrogates it is approximately $\chi^2$ distributed with $2K$ degrees of freedom. A third way to find evidence for an interaction is based on the well-known phase synchronization criterion. This criterion imposes constraints on the difference between integer multiples of the two phases under study [5-7],

$$|n\,\Phi_j - m\,\Psi_j| \;<\; const \qquad (11)$$

where $\Phi_j$ and $\Phi_j$ denote the unlimited (smoothed out) extensions of $\varphi_j$ and $\psi_j$. To account for gliding co-ordination [27] (phase slips), extensions to this strict definition of phase co-ordination have been proposed, which include more general properties of the above phase difference [6,7,11,28]. One of these extensions is based on residence times [29] in the phase coordination mode. Residence times are determined as the average waiting time until condition (11) is violated. For coordinated dynamics the residence time is large in comparison to the larger of the two average cycle lengths. The relative residence time is the ratio to this cycle length.

In contrast to the last criterion the first ones also allow the distinction of the direction of an interaction. If the dynamics of a subsystem is found to be asymptotically stable under the evident influence of the other sub-system, we can speak of conditional asymptotic stability.

Unilateral conditional asymptotic stability of one of two oscillators (the passive one) guarantees that the active oscillator exerts an evident effect on the passive one and that this effect is reproducible irrespective of past and present states of the passive subsystem (within a certain local neighborhood of the reference trajectory). However this effect is dependent on past and/or present states of the active oscillator. This means that there exists a set of states of the active oscillator, which uniquely determines a state of the passive oscillator. In the case of "generalized synchronization in unidirectionally coupled systems" [15] this active history of states degenerates to the simultaneous state of the active oscillator, leading to the well known Afraimovich map relating simultaneous states [13]. Note that asymptotically stable phase synchronization (with n:m ≠ 1:1 coordination) includes cases with a finite length active history.

In the case of bilateral conditional asymptotic stability it is important to note that the passive reference trajectory in one of the two gedanken experiments is identical to the trajectory with the active role in the other gedanken experiment. This means that there exist two unique maps acting in opposite directions, which relate subsets of states of one trajectory to single states of the other. In many cases the two maps will simplify to a single invertible map. To avoid a conflict with the existing notion of phase synchronization, all cases of either unilateral or bilateral conditional asymptotic stability with either zero or finite length history map(s) should fall under the notion of generalized synchronization.

Like normal Liapunov exponents the conditional ones are topological invariants. The equivalence of FIPS map reconstructions for different canonical phases is the basis to detect conditional asymptotic stability of canonical phases as a robust system property independent of the particular choice of the phase. Note that residence times in one particular phase coordination mode are not topologically invariant and known to dependent strongly e.g. also on filters used for the preprocessing.

Prior to the application of the newly developed concepts to cardio-respiratory interaction, a mathematical model is chosen as a test case,

$$X_{n+1} = 20\, e^{0.3\,\xi_n}\, X_n\, e^{-0.001\, X_n} + 6\,\kappa_n + 1$$

$$Y_{n+1} = 23\, e^{0.2\,\zeta_n}\, Y_n\, e^{-0.001\,(CX_n + (1-C)Y_n)} + 4\,\lambda_n + 1$$

(12)

where $\xi_n$, $\zeta_n$ represent independently, (0,1) distributed Gaussian random numbers and $\kappa$, $\lambda$ two corresponding uniform random processes. (A more symmetric version of the model might be considered as useful to study functional aggregation in a two "species" ecosystem of two virulent populations, like (yearly) measles cases [9] in two large neighboring cities.) The phase corresponding to X is obtained as $\psi_n = \arctan 2(\log_{10} X_n - 3,\ \log_{10} X_{n+1} - 3)$ and the second phase $\varphi$ in complete analogy. As can be seen from line one and three of Table II the parameter values are chosen in such a way that the conditional Liapunov exponent of oscillator *y* is negative for *C=0.6* and slightly positive for *C=0.2*. FIPS map reconstructions based on the phenomenological ansatz,

$$\tan(\varphi_{n+1})\sin(\varphi_n) = a_0 + \sum_{k=1}^{4} A_{k,0} \sin(k\,\varphi_n + \alpha_{k,0}) + \sum_{k=2}^{4} O_{k,0} \sin(k\,\psi_n + \omega_{k,0})$$
$$+ C_{1,-1} \sin(\varphi_n - \psi_n + \gamma_{1,-1}) + C_{1,1} \sin(\varphi_n + \psi_n + \gamma_{1,1})$$

(13)

and the symmetric analogon for $\psi$ have been generated for 12 consecutive sections containing 500 pairs of phases each (Fig. 1). To avoid the more involved version of linear regression based on singular value decomposition in the case *C=0.6* (with near identical phase synchronization), the first order open loop terms have been eliminated. As an example Fig. 1 shows one of the estimated FIPS map generating functions for the reconstruction of $\psi$.

The second and third line of Table II indicate averages and standard deviations of FIPS map reconstructed conditional LE, relative residence times and test statistics for nonzero cross impact. The comparatively small standard deviations demonstrate the high degree of reproducability of the FIPS map reconstruction. In view to the arbitrarily chosen reduction factor (1/2) of the noise level used for the reconstruction, the average conditional LE and residence times are in reasonable agreement to the corresponding values directly obtained from the model. The $\chi^2$ statistics (for 10 degrees of freedom) leave no doubt about the presence of cross impact (the 99.9% significance level is about 30). As should be expected the large difference between the test statistics for the two phases expresses the unidirectional character of the coupling.

**Cardio-respiratory interaction**

The empirical example is based on 20 minutes recording of cardio-respiratory data from each of a group of 12 healthy, young adult, male volunteers in relaxed state. The cardiac signal is represented by heartbeat intervals $\{I_n | n = 0, 1, ..., N\}$ and the respiratory signal by airflow $\{F_n = F(t_n) | n = 0, ..., N\}$ sampled at the non-equidistant set of points in time $\{t_n | t_n = t_{n-1} + I_{n-1}; n = 1, ..., N\}$ defined by systolic events. Both signals are subjected to a high-pass filter eliminating the frequency components below 0.1 Hz. This sets the focus on the respiratory sinus arrhythmia at about 0.25 Hz as well as on its first subharmonic(s) (Fig. 2). Cardiac signal, $I$, is transformed into a set of cardio-respiratory phases, $\{\varphi_n | n = 1, ..., N\}$ and $F$ is transformed analogously into respiratory phases $\psi_n$. For each subject, 10 partially overlapping sets of 300 heartbeat lengths and corresponding systolic airflows are selected to estimate a FIPS map according to

$$\tan(\varphi_{n+1})\sin(\varphi_n) = a_0 + \sum_{k=1}^{3}\left[A_{k,0}\sin(k\,\varphi_n + \alpha_{k,0}) + A_{0,k}\sin(k\,\varphi_{n-2} + \alpha_{0,k})\right]$$
$$+ \sum_{k=1}^{3} O_{k,0}\sin(k\,\psi_n + \omega_{k,0}) \quad (14)$$
$$+ C_{1,-1}\sin(\varphi_n - \psi_n + \gamma_{1,-1}) + C_{1,1}\sin(\varphi_n + \psi_n + \gamma_{1,1})$$

and the symmetric respiratory analogon. In all 120 cases the $\chi^2$ statistics for first order open loop, respiration induced cross impact, $O_{1,0}$ is found to be significant at the 99.9% limit (compared to no cases for surrogate data obtained by a time shift of 100 heartbeats between the two cardio-respiratory signals). The corresponding statistics for the open loop cross impact acting *on* the respiration turn out to be significant in 64 cases (at the 95% limit). The FIPS map reconstruction evidences 118 cases with non-trivial (non fixed point) attractors. Their conditional Liapunov exponents $(\lambda_c, \lambda_r)$ reveal 116 cases with conditional asymptotic stability of the cardio-respiratory phase ($\lambda_c \leq -0.1$) and 48 cases of bilateral conditional AS ($\lambda_r \leq -0.05$). The near ubiquitous conditional AS of the cardio-respiratory phase, in particular, has also been found for a truncation of the Fourier series at second order. In 104 cases the deterministic reconstruction evidences a (1:1) phase coordination according to (6) (relative residence time $\geq 10$) and in 12 more cases either (2:3) or (1:2) coordination is encountered. (The latter numbers are obviously dependent on the cut off frequency of the high pass filter.)

In 102 cases the largest LE (of the combined dynamics) is close to zero, indicating quasi-periodic motion and the appearance of a one-dimensional manifold (curve) relating the respiration-related phases. Its two-dimensional projection on the plane of the simultaneous phases can be used for a characterization of the degree of deviation from identical phase synchronization (Fig. 3). The first qualitative step of deviation from a straight line is characterized by non-monotonicities of the $\varphi_n$ vs. $\psi_n$ map (Fig. 3) and the second qualitative step by non-uniqueness of this map indicating an active history of the respiratory phase (n:m ≠ 1:1). Whereas more than half of the subjects evidence monotonous, invertible Afraimovich maps, a few subjects show strong deviations from identical phase synchronization. The non-monotonicities occur at typical respiratory phases (Fig. 3).

Temporarily the conditional asymptotic stability enters a higher state of order characterized by a negative largest LE, a discrete definition set of the asymptotically stable maps and commensurability to a certain number of sampling intervals [6,18]. However, such cases of "phase locking" represent only a minor subset. It is expected that amplitudes of the cross impact on the cardio-respiratory phase constitute more robust and/or specific measures of cardio-respiratory interaction than measures based either on the degree of phase locking [6,18] or the degree of phase coordination [6,28].

The cross impacts between the two respiration related oscillators evidence a clear asymmetry. The near permanent cross impact on the cardio-respiratory phase is contrasted by a (weaker) temporarily active cross impact on the respiratory one. The two oscillators are known to differ in their accessibility to voluntary action [20,30] as well as to conscious self-perception. It is hypothesized that a significant conditional asymptotic stability of the respiratory phase, conditioned on the cardio-respiratory one, can be taken as indicator for involuntary, spontaneous breathing and the evident absence of cross impact from the cardio-respiratory phase on the respiratory one as (objective) indicator for active, voluntary respiration. The role of the astonishingly rich variety of different synchronization pattern should be analyzed by further empirical studies.

**Conclusion**

After more than 150 years of scientific records [19] the "influence of respiratory movement on the blood current" can be described as asymptotic stability of a unique map, which relates the phase of the heartbeat modulation in the frequency range of breathing to phases of the respiratory activity and which often fulfils the (1:1) phase coordination criterion. The uni- or bilateral conditional asymptotic stability (AS) has been identified in reconstructions of cardio-respiratory data of 12 subjects based on finite order Fourier-approximated, Invertibility enforcing Phase Space maps. FIPS map dynamics are shown to be topologically equivalent for a whole set of different canonical phases. In connection with the topological invariance of Liapunov exponents this is the basis to detect conditional AS as a robust system property. Conditional AS includes an evidence for the presence of cross impact. When applied to FIPS map reconstructions of time series it represents a criterion, which unites several non-exclusive synchronization phenomena including different types of (n:m) phase synchronization, generalized synchronization of unidirectionally coupled phases and commensurability to sampling intervals. Apart from being more robust and general, FIPS map based criteria are potentially more specific than existing synchronization or coordination criteria applicable to time series [5-7, 11,15, 17-18,28,30]. Significance tests for nonzero cross impact amplitudes of FIPS maps can be used as a sensitive indicator for the presence of interaction including the possibility to distinguish unidirectional coupling or active – passive relationships. The latter features may e.g. be used to monitor spontaneous breathing. In the more general context of brain science cardio-respiratory interaction may turn out as a prototype system to further the understanding of the two philosophically charged dichotomies, voluntary vs. involuntary action and consciously perceived vs. unconscious brain processes, in terms of FIPS map reconstructed neural dynamics.

I extend my thanks for helpful discussions to M. Schiek, P. Grassberger, H. Halling , P. Tass and H. Müller-Krumbhaar, Jülich, to N. Stollenwerk, Cambridge, UK, R. Engbert, Potsdam, J. Schnakenberg, Aachen and to H.-H. Abel, Braunschweig, in particular for providing the empirical data set.

1. F. Takens, *Lecture Notes in Mathematics* **898**, (Springer, Berlin, 1981)

2. N.H. Packard, J.P. Crutchfield, J.D. Farmer and R.S. Shaw, *Phys. Rev. Lett.* **45**, 712 (1980)

3. H. Kantz, T. Schreiber, Nonlinear time series analysis, Cambridge Univ. Press (1997)

4. D. Gabor, J. IEE London **93**, 429 (1946)

5. M.G. Rosenblum, A.S. Pikovsky, J. Kurths, *Phys. Rev. Lett.* **76**, 1804 (1996)

6. M.G. Rosenblum, J. Kurths, A. Pikovsky, C. Schäfer, P. Tass, H.-H. Abel, *IEEE Eng. Med. Biol.* **17**, 46-53 (1998)

7. P. Tass, M.G. Rosenblum, J. Weule, J. Kurths, A. Pikovsky, J. Volkmann, A. Schnitzler, H.-J. Freund, *Phys. Rev. Lett.* **81,** 3291 (1998)

8. M. Casdagli, Physica D, **35**, 335-356 (1989)

9. F.R. Drepper, R. Engbert, N. Stollenwerk, Ecol. Mod., **75/76**, 171-181 (1994)

10. C. Hugenii, Horoloquium Oscillatorium. *Parisis*, France (1673)

11. R.L. Stratonovich, Topics in the theory of random noise. (Gordon and Breach, New York, 1963)

12. H.Fujisaka & T. Yamada, Prog. Theor. Phys. **69**, 32-47 (1983)

13. V.S. Afraimovich, N.N. Verichev, M.I. Rabinovich, *Radiophys. Quantum Electron.* **29**, 795 (1986)

14. L.M. Pecora and T.L. Caroll, *Phys. Rev. Lett.* **64**, 821 (1990)

15. N.F. Rulkov, M.M. Sushchik, L.S. Tsimring, H.D.I. Abarbanel, *Phys. Rev. E* **51**, 980-994 (1995)

16. L. Kocarev, U. Parlitz, *Phys. Rev. Lett.* **76**, 1816 (1996)

17. S.J. Schiff, P. So, T. Chang, R.E. Burke, T. Sauer, *Phys. Rev. E* **54**, 6708-6724 (1996)

18. C. Schäfer, M.G. Rosenblum, J. Kurths, H.-H. Abel, *Nature* **392**, 239-240 (1998)

19. C. Ludwig, *Arch. Anat. Physiol.* **13**, 242-302 (1847)

20. C.T.M. Davies and J.M.M. Neilson, *J. Appl. Physiol.* **22**, 947-955 (1967)

21. J.A. Hirsch and B. Bishop, *Am. J. Physiol.* **241**, H620-H629 (1981)

22. H.P. Koepchen, In *Mechanisms of Blood Pressure Waves*, edited by K. Miyakawa, H.P. Koepchen and C. Polosa, (Springer, Berlin, 1984)

23. M.P. Gilbey, D. Jordan, D.W. Richter, K.M. Spyer, *J. Physiol.* **365**, 65-78 (1984)

24. D.L. Eckberg, Y.T. Kifle and V.L. Roberts, *J. Physiol.* **304**, 489-502 (1980)

25. K. Suder, F.R. Drepper, M. Schiek, H.-H. Abel, *Am. J. Physiol.* **275**, H1092-H1102 (1998)

26. J. Arnhold, P. Grassberger, K. Lehnertz, C.E. Elger, Physica D **134**, 419-430 (1999)

27. E.v. Holst, *Ergebn. Physiol.* **42**, 228-306 (1939)

28. M. Schiek, RWTH Aachen, Aachen, PhD thesis, 1998

29. A. Neiman, A. Silchenco, V. Anishchenko, L. Schimanski-Geier, Phys. Rev. E **58**, 7118-7125 (1998)

30. M. Schiek, F.R. Drepper, R. Engbert, H.-H. Abel, K. Suder, In *Nonlinear Analysis of Physiological Data*, edited by K. Kantz, J. Kurths and G. Mayer-Kress, (Springer, Berlin, 1998)

TABLE I. Coefficients $\beta$, $\gamma$ and $\delta$ of FIPS maps and canonical phases defined by two linear functions $L_x$ and $L_y$ describing two orthogonal projections in the primary reconstruction space.

| $L_y(x_0,x_1,x_2)$ | $L_x(x_0,x_1,x_2)$ | $\beta$ | $\gamma$ | $\delta$ |
|---|---|---|---|---|
| $\sin\alpha\, X_1 + \cos\alpha\, X_2$ | $\cos\alpha\, X_1 - \sin\alpha\, X_2$ | $\alpha$ | $\alpha$ | $\alpha$ |
| $(X_0 + 2X_1 + X_2)/\sqrt{6}$ | $(X_0 - X_2)/\sqrt{2}$ | $\pi/3$ | $\pi/6$ | $\pi/6$ |
| $(X_0 - 2X_1 + X_2)/\sqrt{6}$ | $(X_0 - X_2)/\sqrt{2}$ | $\pi/3$ | $-5\pi/6$ | $\pi/6$ |

TABLE II. Results of a study of the model given in Eq. (12) which are explained in the text.

|  | conditional Liapunov exponent for | | relative residence time | $\chi^2$ statistics for cross impact on | |
|---|---|---|---|---|---|
|  | $\psi$ | $\varphi$ |  | $\psi$ | $\varphi$ |
| simulation for C=0.6 | 0.28 | -0.29 | 6.0 | | |
| reconstruction | 0.36±0.03 | -0.47±0.03 | 7.4±1.0 | 4.4±2.1 | 125±35 |
| simulation for C=0.2 | 0.28 | 0.12 | 2.8 | | |
| reconstruction | 0.34±0.02 | 0.10±0.04 | 2.9±0.4 | 4.1±2.4 | 410±79 |

FIG. 1. Reconstruction generating function $G(\psi,\varphi)$ as function of $\psi$ for the driver dynamics of model (12) with $c=0.2$. The 500 open circles represent the corresponding input values to the estimation based on (ignorant, symmetric) ansatz (13); and full circles represent the resulting reconstruction. The hardly noticeable "noise" of the deterministic reconstruction results from an extremely weak coupling to the response dynamics - a fact which reflects the near perfect "recognition" of the unidirectional coupling. Phase $\psi$ is given in units of $\pi$.

FIG. 2. High-pass filtered heartbeat interval lengths (full circles) and high-pass filtered respiratory flow signal (open squares) sampled at systolic events 670-730 of subject 9. The heartbeat intervals (relative deviations from an average of 957 ms) are obtained from an ECG recording (1Ksps) and the flow signal is obtained from an uncalibrated thermistor attached to the nose.

FIG. 3. Projection of a cardio-respiratory FIPS map reconstruction (full circles) on the plane defined by the two simultaneous phases and corresponding scatter diagram of the underlying empirical data (300 open circles, partially shown in Fig. 2). This projection reveals an asymptotically stable map, which relates cardio-respiratory phase $\varphi$ uniquely to the simultaneous respiratory phase $\psi$. Both phases are given in units of $\pi$. Subject 9 evidences strong deviations from identical phase synchronization.

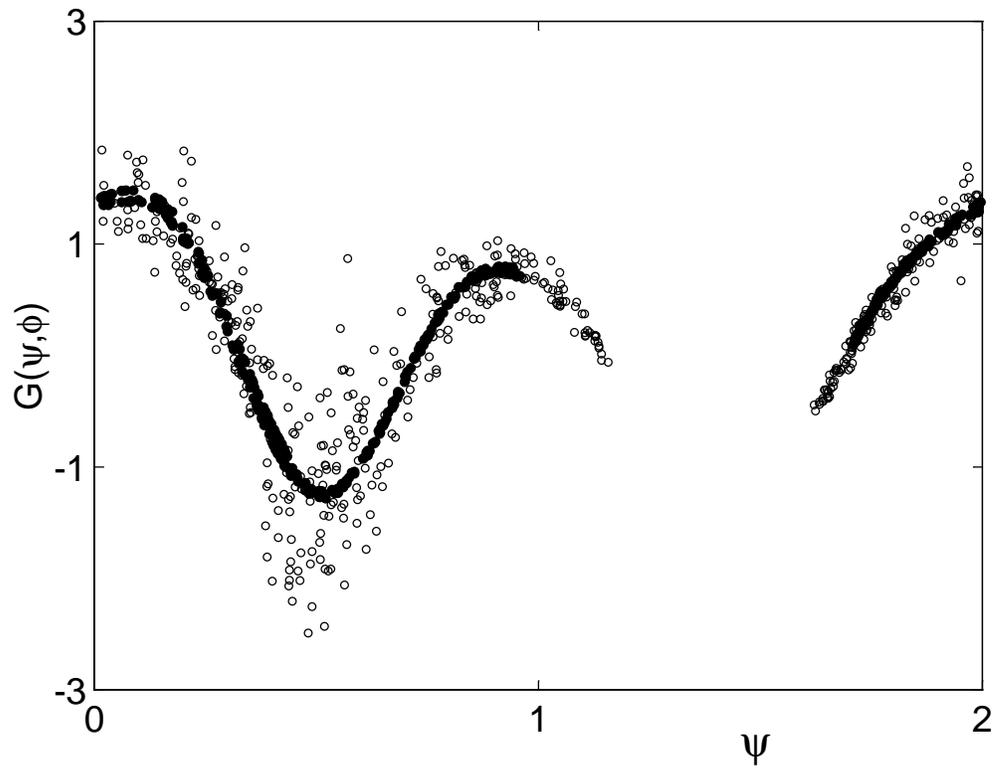

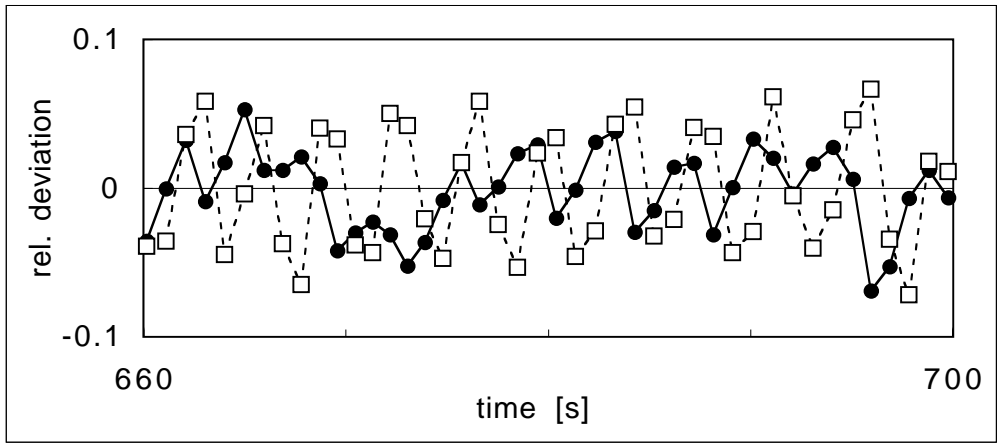

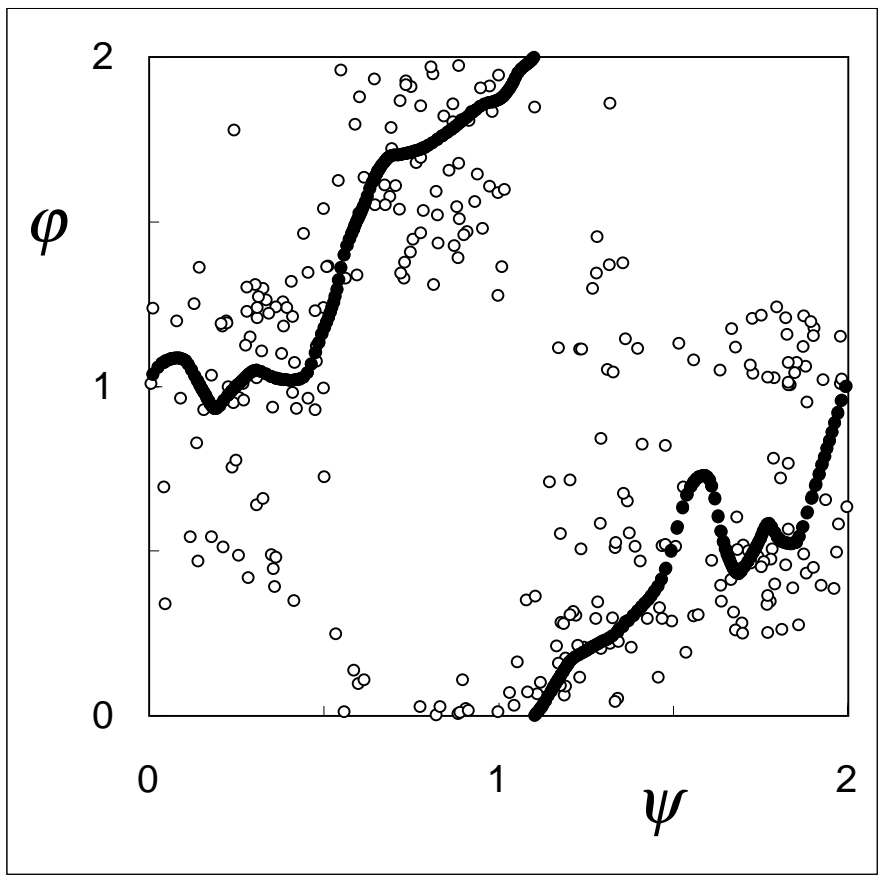